\documentclass[aps,prl,twocolumn,superscriptaddress,nofootinbib]{revtex4-2}
\usepackage[english]{babel}
\usepackage{bm}
\usepackage{amsmath}
\usepackage{amssymb}
\usepackage{textcomp}
\usepackage{dutchcal}
\usepackage{graphicx}
\usepackage[colorlinks=true, allcolors=cyan]{hyperref}
\usepackage{diagbox}
\usepackage{multirow}
\usepackage{threeparttable}
\usepackage{xfrac}
\usepackage{xcolor}

\usepackage{booktabs}

\begin{document}

\title{Seebeck and Hall effects in charge-neutral ballistic graphene as probes of electron-electron interactions}

\author{Z.Z. Alisultanov}
\email{zaur0102@gmail.com}
\affiliation{Moscow Institute of Physics and Technology, Institutskiy per. 9, Dolgoprudny 141701, Russia}
\author{A.V. Kavokin}
\email{kavokinalexey@gmail.com}
\affiliation{International Center for Quantum Optics and
Quantum Technologies, 11, Tsvetnoy Blvd, Moscow, 127051, Russia}
\affiliation{Moscow Institute of Physics and Technology, Institutskiy per. 9, Dolgoprudny 141701, Russia}

\begin{abstract}
Electron-hole asymmetry of the Coulomb velocity renormalization is responsible for finite Seebeck and Hall coefficients in charge-neutral graphene. In the ballistic regime the thermopower is small, $0.1$--$1~\mu$V/K, because the chemical-potential shift required by $n_e{=}n_h$ cancels its leading part. When the electron-electron scattering length falls below the sample length, the thermopower crosses over to a collision-dominated value of $4$--$10~\mu$V/K, that is an order of magnitude larger. This crossover is a direct thermoelectric marker of the breakdown of ballistic transport and a baseline for identifying the Dirac fluid.
\end{abstract}

\maketitle

Graphene is a convenient platform for thermoelectric transport because the Dirac spectrum makes the thermopower a sensitive probe of the carrier type and of the scattering mechanism~\cite{Novoselov2004,CastroNeto2009}. Measurements of the Seebeck coefficient $S$ as a function of the gate voltage show that $S$ changes sign at the charge-neutrality point and it is well described by the Mott relation away from it~\cite{Zuev2009,Wei2009,Checkelsky2009,Hwang2009}. In clean samples at elevated temperatures the Mott relation and the Wiedemann--Franz law are violated owing to inelastic electron--electron scattering and the formation of a Dirac fluid~\cite{Ghahari2016,Crossno2016,Muller2009,Fritz2008,Lucas2018}. Thermoelectric properties of ballistic graphene and related Dirac systems can be conveniently described in the framework of Landauer formalism~\cite{Dragoman2007,Kavokin2024_PRB,Alisultanov2025_PRB_topo}.

In Dirac approximation, the energy spectrum of graphene is electron--hole symmetric, which is why both Seebeck and Hall coefficients vanish at the charge neutrality point. This one-particle result has not been tested experimentally, since absolutely neutral samples do not exist: charge inhomogeneity always shifts the Fermi level from the Dirac point. The situation is changing nowadays: the use of hBN substrates helps reducing the inhomogeneity of samples~\cite{Dean2010,Xue2011}, while the proximity screening in a 1-nm hBN dielectric layer may bring the residual carrier density down to $\delta n\sim3\times10^{7}~\mathrm{cm}^{-2}$~\cite{domaretskiy2025proximity}, with potential fluctuations well below $k_BT$. Experiments in the immediate vicinity of the charge neutrality point manifested by symmetric populations of electron and hole bands thus become feasible.

It should be noted that the Dirac approximation holds only near the Dirac point; the full band structure of graphene is electron--hole asymmetric: the $\pi$ and $\pi^*$ bands have different widths and curvatures because of hopping to next to nearest neighbors~\cite{Wallace1947,Saito1998,Reich2002,CastroNeto2009}. This asymmetry exists already in a single-particle picture, but it becomes observable in transport measurements only due to the interactions. Coulomb interaction renormalizes the Fermi velocity in the leading logarithmic approximation, $v_F(E)=v_0[1+(\alpha/4)\ln(\Lambda/E)]$, where $\alpha=e^2/(\epsilon\hbar v_0)$ and $\Lambda$ is an ultraviolet cutoff of the order of the bandwidth~\cite{GGV1994,GGV1999,Kotov2012}. The logarithmic growth of the velocity toward the Dirac point was observed in cyclotron-mass, quantum-capacitance, and photoemission experiments~\cite{Elias2011,Siegel2011,Yu2013}. Because the cutoffs $\Lambda_+$ and $\Lambda_-$ of the electron and hole bands differ, the renormalization is itself electron--hole asymmetric. Indeed, an asymmetry of the renormalized velocity between electron and hole doping has been observed by photoemission in graphene/hBN devices and attributed to correlation effects~\cite{Dale2022}.

The effect of velocity renormalization on transport in graphene is subtle. In Kubo formalism, the logarithms of the self-energy and the vertex corrections cancel in the conductivity of clean graphene~\cite{Mishchenko2007,Mishchenko2008,Sheehy2007,Herbut2008,Juricic2010}. In the ballistic regime, however, within the Landauer formalism~\cite{SivanImry1986,Datta1995} electric current is dependent on the number of propagating modes, which for a wide strip is governed by the radius $k_F(E)=E/\hbar v_F(E)$ of the constant-energy contour, $M\propto k_F$, rather than by the density of states $\propto1/v_F^2$. Therefore, the number of modes depends on the renormalized velocity, and an electron--hole asymmetry of the cutoffs makes $M_+(E)\neq M_-(E)$. One might expect unequal numbers of electron and hole modes to give a nonzero Seebeck coefficient in neutral graphene. The central issue is the definition of the neutral state. Note that in a gated device the net charge $n_e-n_h$ is fixed, not the chemical potential, and for asymmetric bands $n_e=n_h$ requires a temperature-dependent shift $\mu(T)$ from the Dirac point, which itself contributes to the thermopower.

\begin{figure*}[t]
    \centering
    \includegraphics[width=\textwidth]{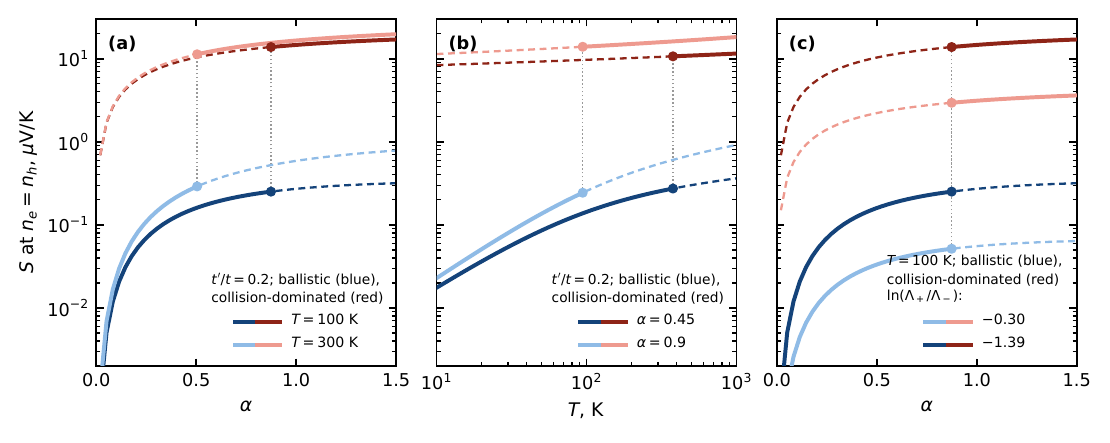}
    \caption{Seebeck coefficient of graphene at charge neutrality, $n_e=n_h$, versus $\alpha$ and $T$ in the two transport regimes. Blue: ballistic regime, Eq.~\eqref{eq:S} with the minimal-conductivity term. Red: collision-dominated regime, $S=-\mu_n/eT$, valid for $\ell_{ee}<L_s$, when local equilibrium is established and the transport function is $\sigma\propto\nu v_g^2\tau\propto E$. Solid lines show the range of validity of each regime, dashed lines its continuation beyond the crossover $\ell_{ee}=L_s$ (dots), i.e.\ $T=T^*=\hbar v_0/\alpha^2k_BL_s$ or $\alpha^2=\hbar v_0/k_BTL_s$; dotted vertical segments connect the values of $S$ in the two regimes at the crossover. Two-color segments in the legends denote the ballistic and collision-dominated curves for the same parameters. $L_s=0.1~\mu$m. (a)~$S$ versus $\alpha$ at $T=100$ and $300$~K, $t'/t=0.2$ ($\ln(\Lambda_+/\Lambda_-)=-1.39$). (b)~$S$ versus $T$ at $\alpha=0.45$ and $0.9$, $t'/t=0.2$. (c)~$S$ versus $\alpha$ at $T=100$~K for two values of the asymmetry, $\ln(\Lambda_+/\Lambda_-)=-0.30$ and $-1.39$ ($t'/t=0.05$ and $0.2$).}
    \label{fig:S}
\end{figure*}

In this Letter we show that the thermopower of graphene at $n_e=n_h$ exhibits two distinct regimes. In the ballistic regime the shift $\mu(T)$ cancels the contribution of the mode asymmetry to the linear order in asymmetry, and the remaining thermopower, $0.1$--$1~\mu$V/K, is governed by the energy dependence of the renormalized velocity. When the electron--electron scattering length $\ell_{ee}$ becomes shorter than the sample length $L_s$, the mode-number contribution disappears and the thermopower crosses over to $-\mu(T)/eT$, that is more than an order of magnitude larger (Fig.~\ref{fig:S}). This crossover can be considered as a thermoelectric marker of the breakdown of ballistic transport. It characterises the temperature range in which the ballistic transport can be observed. We also show that, unlike the thermopower, the semiclassical Hall coefficient remains linearly dependent on the electron--hole asymmetry at the charge neutrality condition.

\begin{figure}[]
\centering
\includegraphics[width=\linewidth]{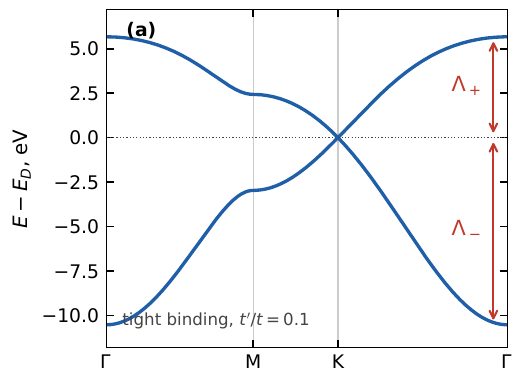}
\includegraphics[width=\linewidth]{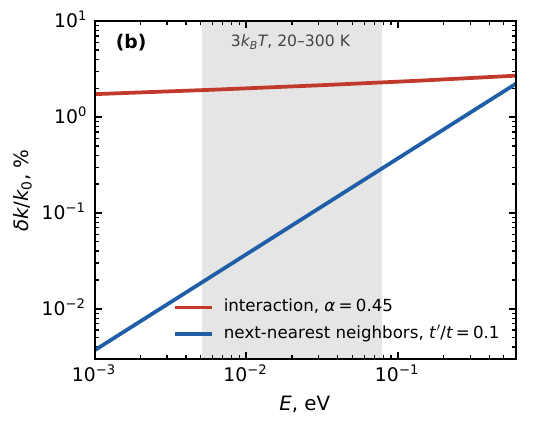}
\caption{Two sources of electron--hole asymmetry. (a)~Tight-binding bands with next-nearest-neighbor hopping, $t'/t=0.1$; $\Lambda_\pm$ are the distances from the Dirac point to the band edges at $\Gamma$. (b)~Relative asymmetry $\delta k/k_0=(k_+-k_-)/(k_++k_-)$ of the radius of the constant-energy contour: for the interaction ($\alpha=0.45$) it is nearly independent of energy, for next-nearest-neighbor hopping it grows linearly with $E$. Shading: $3k_B T$ for $T=20$--$300$~K.}
\label{fig:mech}
\end{figure}

\emph{Model.}---Following the leading-logarithm result for the Coulomb interaction~\cite{GGV1994,GGV1999,Kotov2012} we write the renormalized velocity of electrons and holes as
\begin{equation}
v_\pm(E)=v_0\Bigl[1+a\ln\frac{\Lambda_\pm}{E}\Bigr],\quad a=\frac{\alpha}{4},\quad \Lambda_+\neq\Lambda_-,
\label{eq:v}
\end{equation}
where $E>0$ is the energy of an electron in the $\pi^*$ band or the absolute value of the energy of a hole in the $\pi$ band, measured starting from the Dirac point, and $E<\Lambda_\pm$. Here the constant $\alpha$ is not assumed to be small: we consider $0<\alpha<3/2$, where the one-loop renormalization is the leading effect of perturbation theory~\cite{Mishchenko2007}. The asymmetry is characterized by
\begin{equation}
L\equiv\ln\frac{\Lambda_+}{\Lambda_-}.
\label{eq:L}
\end{equation}
 $\Lambda_\pm$ denote the energy distances from the Dirac point to the band edges at $\Gamma$ [Fig.~\ref{fig:mech}(a)]. In the tight-binding model with next-nearest-neighbor hopping $t'$ they are $\Lambda_\pm=3t\mp9t'$, so $L=\ln[(1-3t'/t)/(1+3t'/t)]$, and $L=-0.62$ for $t'/t=0.1$ ($t=2.7$~eV)~\cite{CastroNeto2009,katsnelson2020physics}.

\emph{Ballistic thermopower.}---For a strip of width $W$ between ideal contacts the conductance is $G=(ge^2/h)\int dE\,w(E-\mu)\sum_n\mathcal T_n(E)$, where $\mathcal T_n(E)$ is the transmission probability of the $n$-th propagating mode at the energy $E$, $g=4$, and $w(x)=-f'(x)=[4T\cosh^2(x/2T)]^{-1}$ ($k_B=1$ unless stated otherwise). According to the Landauer ballistic formalism the Seebeck coefficient is:
\begin{equation}
S=-\frac{1}{eT}\,
\frac{\int_{-\infty}^{\infty} dE\,(E-\mu)\, M(E)\, w(E-\mu)}
     {\int_{-\infty}^{\infty} dE\,M(E)\, w(E-\mu)+M_{\min}},
\label{eq:S}
\end{equation}
where $e>0$, $M$ is the number of propagating modes per spin-valley copy, and $M_{\min}$ accounts for the minimal conductivity of graphene. We assume unit transmission of every propagating mode; the limits of this assumption, set by electron--electron scattering, are discussed below in connection with the crossover.

The transverse momentum in a wide strip is quantized, $q_n=2\pi n/W$ (periodic) or $q_n=\pi n/W$ (hard wall), and a mode contributes to the transport if $|q_n|<k_F(E)$. In both cases the number of modes is $Wk_F/\pi$, which is why
\begin{equation}
M(E)=\frac{W}{\pi}\,k_{\text{sgn} E}(|E|),\quad k_\pm(E)=\frac{E}{\hbar v_\pm(E)}.
\label{eq:M}
\end{equation}
The number of propagating modes thus depends on the position of the constant-energy contour in momentum space, but not on its slope. The group velocity only affects the density of states, $\nu_\pm=(g/4\pi)\,dk_\pm^2/dE$, used below to define the neutrality condition. This is specific to the ballistic transport regime. In contrast, in the diffusive (Kubo) regime the conductivity is governed by $v_g^2\,\nu(E)$, in which the dependence on the velocity cancels. Writing $k_\pm=k_0\pm\delta k$, one finds exactly
\begin{gather}
k_0=\frac{E}{\hbar v_0}\,\frac{u}{u^2-d^2},\quad
\delta k=-\frac{E}{\hbar v_0}\,\frac{d}{u^2-d^2},\nonumber\\
\frac{\delta k}{k_0}=-\frac{d}{u},
\label{eq:k0dk}
\end{gather}
with $u=1+a\ln(\bar\Lambda/E)$, $d=aL/2$, and $\bar\Lambda=\sqrt{\Lambda_+\Lambda_-}$. The symmetric part $k_0$ is a nonlinear function of $E$ because of the logarithm, while $\delta k/k_0$ is nearly energy independent [Fig.~\ref{fig:mech}(b)].

For a sample of length $L_s$ the minimal conductivity $\sigma_{\min}=4e^2/\pi h$ corresponds to the conductance $G_{\min}=\sigma_{\min}W/L_s$, that is, to
\begin{equation}
M_{\min}=\frac{W}{\pi L_s}
\end{equation}
in the units of Eq.~\eqref{eq:S}. It is energy independent and symmetric in $E\to-E$, which is why it does not contribute to the numerator, $\int dE\,(E-\mu)\,w(E-\mu)=0$. Its weight in the denominator is governed by the dimensionless parameter $\eta=\hbar v_0/(k_BTL_s)$; the two terms of the denominator become comparable at the temperature $T_0=\hbar v_0/(2\ln2\,k_BL_s)$ ($T_0\approx5.5$~K for $L_s=1~\mu$m and $v_0=10^6$~m/s).

\emph{Neutrality condition.}---The $\pi$ and $\pi^*$ bands contain the same number of states, because the size of the Brillouin zone does not depend on the band widths. Charge neutrality therefore implies $n_e=n_h$. At $T=0$ the Fermi level is at the Dirac point, while at $T>0$ and asymmetric bands the chemical potential is shifted to $\mu_n(T)$. With $n_e=\frac{g}{4\pi}\int_0^\infty(k_+^2)'f(E-\mu)\,dE$ and $n_h=\frac{g}{4\pi}\int_0^\infty(k_-^2)'f(E+\mu)\,dE$, integration by parts yields
\begin{equation}
\int_0^\infty dE\,k_+^2(E)\,w(E-\mu_n)=\int_0^\infty dE\,k_-^2(E)\,w(E+\mu_n).
\label{eq:neutral}
\end{equation}
The shift $\mu_n$ enters Eq.~\eqref{eq:S} together with the mode asymmetry. At $\mu=0$ the inequality $M_+\neq M_-$ alone would yield a finite thermopower; at $\mu=\mu_n$ this contribution is cancelled to linear order in the asymmetry, and the surviving thermopower is due to the nonlinearity of $k_0(E)$ produced by the logarithm.

The blue curves in Fig.~\ref{fig:S} show $S$ at $n_e=n_h$ computed from Eq.~\eqref{eq:S} for $0<\alpha<3/2$. The thermopower vanishes at $\alpha=0$ (symmetric bands), grows rapidly at small $\alpha$ (numerically, the exact curves behave as $\alpha^{1.9}$ at $\alpha\lesssim0.1$, in contrast with the linear dependence of the thermopower at $\mu=0$), and bends over to a slower growth at $\alpha\gtrsim1$. At fixed $\alpha$ the thermopower is nearly proportional to $|L|$ (for $\alpha=1$ it is $0.13$, $0.27$, $0.64~\mu$V/K for $L=-0.30$, $-0.62$, $-1.39$), and its sign is opposite to the sign of $L$: for $t'>0$ ($\Lambda_+<\Lambda_-$) the thermopower is positive, hole-like. Representative values at $T=100$~K $\gg T_0$ are given in Table~\ref{tab:S}. In the realistic range the thermopower is of the order of $0.1~\mu$V/K; the largest value ($\approx0.9~\mu$V/K) is reached for $t'/t=0.2$ and $\alpha=1.5$, at the border of applicability of the one-loop renormalization.

\begin{table}[t!]
\caption{Ballistic Seebeck coefficient at $n_e=n_h$ in $\mu$V/K at $T=100$~K $\gg T_0$ for the asymmetry induced by next-nearest-neighbor hopping $t'$ ($t=2.7$~eV).}
\centering
\begin{tabular}{lccc}
\toprule
$t'/t$ & $0.05$ & $0.10$ & $0.20$\\
$\ln(\Lambda_+/\Lambda_-)$ & $-0.30$ & $-0.62$ & $-1.39$\\
\midrule
$\alpha=0.45$ & $0.058$ & $0.119$ & $0.275$\\
$\alpha=1.0$ & $0.133$ & $0.274$ & $0.642$\\
$\alpha=1.5$ & $0.180$ & $0.372$ & $0.878$ \\
\bottomrule
\end{tabular}
\label{tab:S}
\end{table}

At $T\gg T_0$ the temperature enters only through $\ln(\bar\Lambda/k_BT)$, so the saturated thermopower depends weakly on $T$. At lower temperatures the minimal conductivity enters through $\eta\propto1/T$: for $T\ll T_0$ the term $M_{\min}$ dominates the denominator of Eq.~\eqref{eq:S}, and $S\propto T$ up to logarithmic factors, so that $S\to0$ at $T\to0$ [Fig.~\ref{fig:S}(b)].

The shift $\mu_n$ is an equilibrium quantity. For $\alpha=0.45$ and $t'/t=0.1$ it is $\mu_n=-0.43$~meV at $T=100$~K, that is, $\mu_n/k_BT\approx-0.05$ with a weak (logarithmic) temperature dependence, so $\mu_n$ is close to linear in $T$. The corresponding net charge density $10^9~\mathrm{cm}^{-2}$ at $100$~K, two orders of magnitude larger than the residual carrier density in the cleanest graphene samples available~\cite{domaretskiy2025proximity}, so the shift $\mu_n$ is in principle experimentally resolvable.

\emph{Interaction versus band asymmetry.}---To compare with the interaction mechanism we consider the asymmetry generated by next-nearest-neighbor hopping without interaction. Near the $K$ point, after angular averaging, $E_\pm(q)=\pm\hbar v_0q-\beta q^2$ with $\beta/(\hbar v_0)^2=t'/t^2$. Inverting exactly,
\begin{equation}
k_\pm=\frac{\hbar v_0}{2\beta}\Bigl[\pm1\mp\sqrt{1\mp\tfrac{4\beta E}{(\hbar v_0)^2}}\Bigr].
\end{equation}
Equations~\eqref{eq:S} and~\eqref{eq:neutral} remain exact with these $k_\pm$. In the dimensionless variables $x=E/k_BT$ and $\kappa_\pm=\hbar v_0k_\pm/k_BT$ the parameter $p=(t'/t^2)k_BT$ is physically small ($p\lesssim10^{-3}$) and allows an analytic expansion, $\kappa_\pm=x\sum_nC_n(\pm px)^n$, where $C_n$ are Catalan numbers. Here $k_0$ is linear in $E$ to $O(p)$ and its nonlinearity appears only at $O(p^2)$, so the first-order terms cancel; the second-order terms vanish as well because $S$ and $\mu_n$ are odd in $t'$. Solving Eq.~\eqref{eq:neutral} order by order gives
\begin{equation}
S_{t'}=c_t\Bigl(\frac{t'k_B T}{t^2}\Bigr)^3\frac{k_B}{e}\,\frac{T}{T+T_0},
\label{eq:Stp}
\end{equation}
with $c_t \approx157.65$. For $t'/t=0.1$ and $T=100$~K this gives $S_{t'}\approx4\times10^{-7}~\mu$V/K and $\mu_n=-0.021$~meV, to be compared with $0.12~\mu$V/K and $-0.43$~meV found above for the interaction mechanism at $\alpha=0.45$. The interaction mechanism is thus larger by five to six orders of magnitude in $S$ and by a factor of about $20$ in $\mu_n$. The difference comes from the nonlinearity of $k_0$, which for interactions is produced by the logarithm and has no small parameter $k_BT/t$, whereas for the band asymmetry it is suppressed as $(t'k_BT/t^2)^2$. The temperature scaling also differs: $S_{t'}\propto T^3$ at $T\gg T_0$ and $\mu_n\propto T^2$, against the nearly temperature-independent $S$ and nearly linear $\mu_n$ for interactions.

\emph{Crossover to the collision-dominated regime.}---The results above assume that every propagating mode is transmitted ballistically. This requires the inelastic electron--electron scattering length $\ell_{ee}\sim\hbar v_0/(\alpha^2k_BT)$ (up to numerical factors~\cite{Fritz2008,Muller2009}) to exceed $L_s$, i.e.\ $T<T^*\equiv\hbar v_0/(\alpha^2k_BL_s)$, while the saturated value of $S$ is reached only for $T\gg T_0$. The window $T_0\ll T\ll T^*$ defines the pure ballistic regime (solid blue curves in Fig.~\ref{fig:S}). Since $T^*/T_0=2\ln2/\alpha^2$, the window is only a factor of about $7$ wide for $\alpha=0.45$ and closes at $\alpha\approx1.2$. At fixed $T$ the crossover occurs at $\alpha^*=(\hbar v_0/k_BTL_s)^{1/2}$, so it can be driven by temperature, sample length, or dielectric screening.

For $T>T^*$ the imaginary part of the self-energy can no longer be neglected. Local equilibrium is established, and transport is described by Boltzmann kinetics. The transport function $\sigma(E)\propto\nu v_g^2\tau\propto E$ then becomes the same for electrons and holes to the leading order, the impact of the unequal numbers of propagating modes vanished, so that nothing compensates the shift of the chemical potential. The thermopower crosses over to $S=-\mu_n/eT$ (red curves in Fig.~\ref{fig:S}), which is $4$--$10~\mu$V/K for $\alpha=0.45$ and $T=100$~K, depending on the asymmetry. This value is more than an order of magnitude larger than the ballistic one, has the same (hole-like) sign for $t'>0$, and, since $\mu_n/k_BT$ is nearly constant, is nearly temperature independent. A measurement of $S(T)$ on a clean sample should therefore show growth $\propto T$ below $T_0$, a plateau of $0.1$--$1~\mu$V/K in the ballistic window, and a rise by more than an order of magnitude at $T^*\propto1/(\alpha^2L_s)$. We model the crossover as a step at $\ell_{ee}=L_s$ (dots in Fig.~\ref{fig:S}); in a real sample it is smeared over a finite range around $T^*$, whose description requires the kinetic equation with electron--electron collisions and lies beyond our scope. Our analysis fixes the two limiting values and the position of the crossover. Since $S=-\mu_n/eT$ in the collision-dominated regime, the ratio of the interaction-induced to the band-asymmetry-induced thermopower falls there to about $20$ (see above), so the ballistic window is where $S$ discriminates most strongly between the two mechanisms.

At still higher temperatures the system enters the hydrodynamic regime, where strong electron--hole scattering near the neutrality condition suppresses the electric current while leaving the carrier heat transport largely unaffected, which leads to an anomalous growth of the Seebeck coefficient~\cite{atlasov2026thermopower}; the Nernst signal is likewise significantly larger~\cite{Muller2008}. The ballistic and Boltzmann values derived here thus serve as reference points to mark the onset of hydrodynamic transport.

\emph{Hall coefficient.}---The Hall effect is determined by the net charge $n_e-n_h$ only in the strong magnetic field limit, $\omega_c\tau\gg1$, where $\rho_{xy}\to B/[e(n_h-n_e)]$. At weak fields, the Hall coefficient of a compensated system depends on the partial conductivities of electrons and holes, and in our system it does not vanish at $n_e=n_h$ because of the electron--hole asymmetry. Geometric oscillations of local Hall and Nernst effects of ballistic graphene in weak fields have been predicted recently~\cite{Alisultanov2026_PRB_geo}. Here we calculate the global Hall coefficient in the semiclassical Boltzmann approximation, which lies outside the local Landauer scheme but relies on the same spectrum $k_\pm(E)$, Eq.~\eqref{eq:M}, and the neutrality condition~\eqref{eq:neutral}.

We consider isotropic bands, assume a weak magnetic field regime $\hbar\omega_c\ll k_B T$, and introduce the relaxation times as $\tau_\pm(E)$. The group velocity is $v_g=\hbar^{-1}dE/dk=v_0/\kappa'$ with $\kappa'=d\kappa/dx$, the cyclotron frequency is $\omega_c=eBv_g/(\hbar k)$, and the density of states is $\nu=gk/(2\pi\hbar v_g)$. To first order in $B$, the partial conductivities of electrons (upper sign) and holes (lower sign) are
\begin{gather}
\sigma_{xx}^{e,h}=\frac{ge^2}{4\pi\hbar}\int_0^\infty dE\,w(E\mp\mu)\,k_\pm v_g\tau_\pm,\\
\sigma_{xy}^{e,h}=\mp\frac{ge^3B}{4\pi\hbar^2}\int_0^\infty dE\,w(E\mp\mu)\, v_g^2\tau_\pm^2,
\end{gather}
and $R_H=\sigma_{xy}/(B\sigma_{xx}^2)$ with $\sigma_{xx}=\sigma_{xx}^e+\sigma_{xx}^h$, $\sigma_{xy}=\sigma_{xy}^e+\sigma_{xy}^h$. We use the dimensionless variables $x=E/k_BT$, $\kappa_\pm=\hbar v_0k_\pm/k_BT$, $W_0(x)=[4\cosh^2(x/2)]^{-1}$, $\kappa'_\pm=(u_\pm+a)/u_\pm^2$, $u_\pm=1+a(X_\pm-\ln x)$, $X_\pm=\ln(\Lambda_\pm/k_BT)$, and $\tilde\tau_\pm(x)=\tau_\pm/\tau_0$. The Hall coefficient at $n_e=n_h=n$ is then
\begin{equation}
R_H e n=\frac{N (I_h-I_e)}{(J_e+J_h)^2}
\label{eq:RH}
\end{equation}
with
\begin{gather}
N=\int_0^\infty\kappa_+^2\,W_0(x-m)\,dx,\quad n=\frac{g}{4\pi}\Bigl(\frac{k_B T}{\hbar v_0}\Bigr)^2 N,\\
I_{e,h}=\int_0^\infty\!\frac{\tilde\tau_\pm^2}{\kappa_\pm'^2}\,W_0(x\mp m)\,dx,\\
J_{e,h}=\int_0^\infty\frac{\kappa_\pm\tilde\tau_\pm}{\kappa_\pm'}\,W_0(x\mp m)\,dx,
\end{gather}
and $m=\mu_n/k_B T$. The scaling parameter $\tau_0$ cancels in Eq.~\eqref{eq:RH}. For a single carrier type Eq.~\eqref{eq:RH} yilds $R_He\,n_{e,h}=\pm1$; for symmetric bands ($\Lambda_+=\Lambda_-$, $m=0$) $I_h=I_e$ and $R_H=0$.

We consider two scattering models with the same approach to electrons and holes: (i) $\tau=\text{const}$, $\tilde\tau=1$; (ii) constant mean free path $\ell=v_g\tau$, $\tilde\tau=\kappa'$, for which $I_{e,h}=\int_0^\infty W_0(x\mp m)\,dx$ and $J_{e,h}=\int_0^\infty\kappa_\pm W_0(x\mp m)\,dx$. The model of short-range scatterers, $\tau\propto1/\nu$, is not suitable at the neutrality point because $\sigma_{xy}$ diverges at $E\to0$.

Table~\ref{tab:hall} shows $R_He\,n$ at $T=100$~K for $t'/t=0.1$ ($L=-0.62$) together with the electron--hole asymmetry of the partial conductivities. The Hall coefficient at the neutrality point is nonzero and it has the positive sign ($R_H>0$) for $\Lambda_+<\Lambda_-$. It grows nearly linearly with $\alpha$ at small $\alpha$ and is weakly temperature dependent (see Table~\ref{tab:hall}). Unlike the thermopower, which cancels to the linear order in asymmetry, the Hall coefficient has no such protection: it remains dependent on the asymmetry of the partial conductivities and of the cyclotron response in first order. The linear-in-B expansion requires $B\ll k_BT/(ev_0^2\tau)$, i.e., a few mT at 100 K for $\tau=1$~ps; Landau quantization is negligible for $B\ll(k_BT)^2/(\hbar ev_0^2)\approx0.11$~T at 100 K. At $T=100$~K and $\alpha=0.45$ one finds $n=3.3\times10^9~\mathrm{cm}^{-2}$ and $R_H=12$~k$\Omega$/T. The bare asymmetry ($t'/t=0.1$) leads to $R_He\,n=0.0026$, about $20$ times smaller. Normalized to its single-carrier value, the Hall response is from one to two orders of magnitude larger than the thermopower, but it is less specific a measurement as it may also be sensitive to puddles and contact doping.

\begin{table}[t!]
\caption{Hall coefficient $R_He\,n$ at $n_e=n_h=n$ and electron--hole asymmetry of the longitudinal partial conductivities, $T=100$~K, $t'/t=0.1$, for two scattering models.}
\centering
\begin{tabular}{lcc}
\toprule
 & \multicolumn{1}{c}{(i) $\tau=$ const} & \multicolumn{1}{c}{(ii) $\ell=$ const}\\
$\alpha$ & $R_He\,n$ & $R_He\,n$\\
\midrule
$0.2$ & $0.035$ & $0.013$ \\
$0.45$ & $0.065$ & $0.022$ \\
$0.9$ & $0.098$  & $0.033$ \\
$1.5$ & $0.124$ & $0.041$ \\
\bottomrule
\end{tabular}
\label{tab:hall}
\end{table}

\emph{Conclusions.}---We have shown that the Coulomb renormalization of the Fermi velocity with different electron and hole cutoffs makes the thermopower and the Hall coefficients of charge-neutral graphene differ from zero at the charge neutrality point. In the ballistic regime, the thermopower of $0.1$--$1~\mu$V/K is driven by the logarithmic energy dependence of the velocity and it exceeds the band-asymmetry induced value by five to six orders of magnitude. Our main finding is the crossover from the ballistic to hydrodynamic regime at $T^*=\hbar v_0/(\alpha^2k_BL_s)$ that is manifested in the strong increase of thermopower. Once $\ell_{ee}<L_s$, local equilibrium is established, the mode-number contribution disappears, and $S$ rises to $-\mu_n/eT=4$--$10~\mu$V/K, that is more than an order of magnitude above the ballistic value. The position of this crossover, $T^*\propto1/(\alpha^2L_s)$, and the change of $S$ across it are a direct thermoelectric marker of the breakdown of ballistic transport and a reference for identifying the hydrodynamic transport regime in ultra-clean graphene. 

\begin{acknowledgments}
The authors thank Timur Pryadilin for a useful discussion of the work. The study was supported by the Ministry of Science and Higher Education of the Russian Federation (State assignment No.~FSMG-2026-0012).
\end{acknowledgments}

\bibliography{apssamp}
\end{document}